\documentclass[preprint]{aastex62}
\usepackage{verbatim}

\begin{document}

\title{White Paper: Exoplanetary Microlensing from the Ground in the 2020s}

\author{Jennifer C. Yee}
\altaffiliation{Primary author; jyee@cfa.harvard.edu; 617-495-7594}
\affiliation{Harvard-Smithsonian Center for Astrophysics, 60 Garden St. MS-15, Cambridge, MA 02138}

\nocollaboration

\author{Jay Anderson}
\affiliation{Space Telescope Science Institute, 3700 San Martin Drive, Baltimore, MD, 21218, USA}

\author{Rachel Akeson}
\affiliation{IPAC, Mail Code 100-22, Caltech, 1200 E. California Blvd., Pasadena, CA 91125, USA}

\author{Etienne Bachelet}
\affiliation{Las Cumbres Observatory, 6740 Cortona Drive, Suite 102, Goleta, CA 93117, USA}

\author{Charles Beichman}
\affiliation{IPAC, Mail Code 100-22, Caltech, 1200 E. California Blvd., Pasadena, CA 91125, USA}

\author{Andrea Bellini}
\affiliation{Space Telescope Science Institute, 3700 San Martin Drive, Baltimore, MD, 21218, USA}

\author{David Bennett}
\affiliation{Code 667, NASA Goddard Space Flight Center, Greenbelt, MD 20771, USA}

\author{Aparna Bhattacharya}
\affiliation{Code 667, NASA Goddard Space Flight Center, Greenbelt, MD 20771, USA}
\affiliation{Department of Astronomy, University of Maryland, College Park, Maryland, USA}

\author{Valerio Bozza}
\affiliation{Dipartimento di Fisica ``E.R. Caianiello", Universit\`{a} di Salerno, Via Giovanni Paolo II 132, 84084, Fisciano, Italy}
\affiliation{Istituto Nazionale di Fisica Nucleare, Sezione di Napoli, Italy}

\author{Sebastiano Calchi Novati}
\affiliation{IPAC, Mail Code 100-22, Caltech, 1200 E. California Blvd., Pasadena, CA 91125, USA}

\author{Will Clarkson}
\affiliation{Department of Natural Sciences, University of Michigan-Dearborn, 4901 Evergreen Road, Dearborn, MI 48128}

\author{David R. Ciardi}
\affiliation{IPAC/NExScI, Mail Code 100-22, Caltech, 1200 E. California Blvd., Pasadena, CA 91125, USA}

\author{Andrew Gould}
\affiliation{Max-Planck-Institute for Astronomy, K\"{o}nigstuhl 17, 69117 Heidelberg, Germany}
\affiliation{Korea Astronomy and Space Science Institute, Daejon 34055, Republic of Korea}
\affiliation{Ohio State University, 140 W 18th Ave, Columbus, OH, USA}

\author{Calen B. Henderson}
\affiliation{IPAC/NExScI, Mail Code 100-22, Caltech, 1200 E. California Blvd., Pasadena, CA 91125, USA}

\author{Savannah R. Jacklin}
\affiliation{Department of Physics and Astronomy, Vanderbilt University, VU Station 1807, Nashville, TN 37235, USA}

\author{Somayeh Khakpash}
\affiliation{Lehigh University}

\author{Shude Mao}
\affiliation{Tsinghua Center for Astrophysics and Department of Physics, Tsinghua University, Beijing, China 100084}

\author{Bertrand Mennesson}
\affiliation{Jet Propulsion Laboratory, California Institute of Technology, 4800 Oak Grove Drive, Pasadena, CA 91109, USA}

\author{David M. Nataf}
\altaffiliation{Allan C. and Dorothy H. Davis Fellow}
\affiliation{Center for Astrophysical Sciences and Department of Physics and Astronomy, The Johns Hopkins University, Baltimore, MD 21218, USA}

\author{Matthew Penny}
\affiliation{Ohio State University, 140 W 18th Ave, Columbus, OH, USA}

\author{Joshua Pepper}
\affiliation{Lehigh University}

\author{Radek Poleski}
\affiliation{Ohio State University, 140 W 18th Ave, Columbus, OH, USA}

\author{Cl\'{e}ment Ranc}
\altaffiliation{NASA Postdoctoral Program Fellow}
\affiliation{Code 667, NASA Goddard Space Flight Center, Greenbelt, MD 20771, USA}

\author{Kailash Sahu}
\affiliation{Space Telescope Science Institute, 3700 San Martin Drive, Baltimore, MD, 21218, USA}

\author{Y.~Shvartzvald}
\altaffiliation{NASA Postdoctoral Program Fellow}
\affiliation{Jet Propulsion Laboratory, California Institute of Technology, 4800 Oak Grove Drive, Pasadena, CA 91109, USA}

\author{R.A.~Street}
\affiliation{Las Cumbres Observatory, 6740 Cortona Drive, Suite 102, Goleta, CA 93117, USA}

\author{Takahiro Sumi}
\affiliation{Department of Earth and Space Science, Graduate School of Science, Osaka University, 1-1 Machikaneyama-cho, Toyonaka, Osaka 560-0043, Japan}

\author{Daisuke Suzuki}
\affiliation{Institute of Space and Astronautical Science, Japan Aerospace Exploration Agency, 3-1-1 Yoshinodai, Chuo, Sagamihara, Kanagawa 252-5210, Japan}

\NewPageAfterKeywords

\section{Overview}

Microlensing can access planet populations that no other
method can probe: cold wide-orbit planets beyond the snow line,
planets in both the Galactic bulge and disk, and free floating planets
(FFPs).  The demographics of each population will provide
unique constraints on planet formation.

Over the past 5 years, U.S. microlensing
campaigns with {\em Spitzer} and {\em UKIRT} have provided a powerful
complement to international ground-based microlensing surveys, with major breakthroughs in parallax measurements and probing new
regions of the Galaxy. The scientific vitality of these projects has
also promoted the development of the U.S. microlensing community.

In the 2020s, the U.S. can continue to play a major role in
ground-based microlensing by leveraging U.S. assets to complement
ongoing ground-based international surveys. {\em LSST} and {\em UKIRT} microlensing surveys
would probe vast regions of the Galaxy, where planets form under
drastically different conditions.  Moreover, while ground-based
surveys will measure the planet mass-ratio function beyond the snow
line, adaptive optics (AO) observations with ELTs would turn all of these
mass ratios into masses and also distinguish between very
wide-orbit planets and genuine FFPs. To the extent possible, cooperation of U.S. scientists with international surveys should also be encouraged and supported.

\section{The Galactic Distribution of Planets}

\begin{figure}[ht!]
\includegraphics[width=3in]{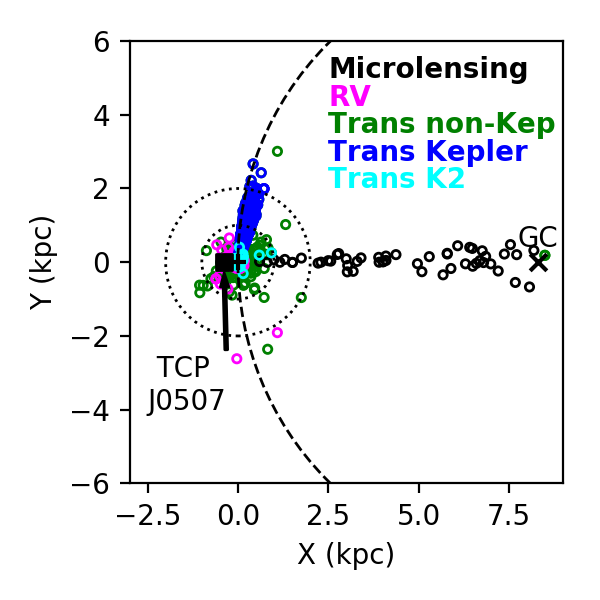}
\includegraphics[width=3in]{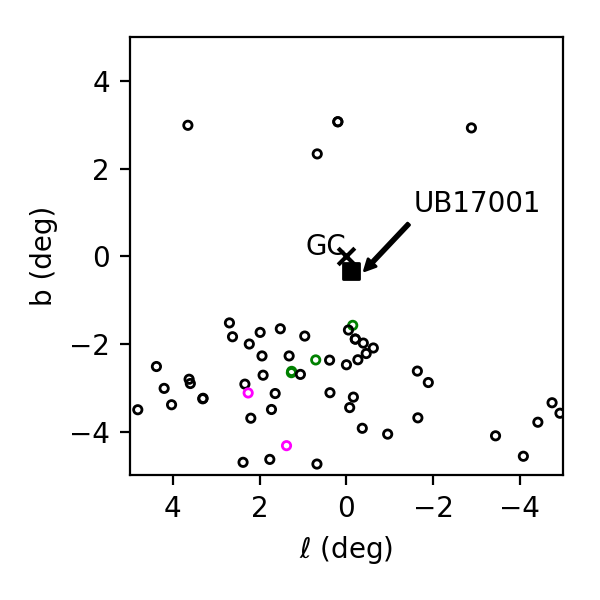}
\caption{The Galactic distribution of known planets found by various
  techniques (circles). Microlensing is the only technique
  that can routinely find planets at distances of more than a few kpc (Left: dashed arc = solar
  circle, dotted circles = 1, 2 kpc from the Sun). TCP
  J05074264+2447555 (TCP J0507) is a microlensing event discovered in
  an all-sky transient survey that has a planet \citep{Nucita18}. {\em
    LSST} observations of the Galactic plane would enable the
  discovery of planets over $\ell \sim (-90, 90)$ degrees (i.e., $X >
  0.$). Right: UKIRT-2017-BLG-001Lb \citep[UB17001;][]{Shvartzvald18}
  demonstrates NIR microlensing surveys can reach
  closer to the Galactic center (x).}
\label{fig:galdist}
\end{figure}

Recent discoveries demonstrate that microlensing planet
searches can reach new areas of the Galaxy outside the traditional
Galactic bulge fields. Figure \ref{fig:galdist} shows that
microlensing is the only technique that probes planets on kpc
scales. An adjusted {\em LSST} observing strategy that includes
the Galactic plane and a continuation of the {\em UKIRT} microlensing
survey, would measure variations in planet frequency on a Galactic
scale. This would provide a powerful probe of how the star-formation environment (e.g., Galactic disk vs. bulge) affects planet formation.

Comparing the planet frequency in the disk and bulge would benefit from three things:
\begin{itemize}
  \item{better distance estimates for microlensing events, i.e., through space-based parallax measurements or AO (see below),}
  \item{more bulge planets, i.e., by observing a large number of far-disk sources through the bulge, which requires high-cadence IR surveys at  $b \sim 0$,}
  \item{finding guaranteed disk planets, i.e., from disk-disk lensing from {\em LSST} plane surveys.}
\end{itemize}

\subsection{Galactic Plane Microlensing with {\em LSST}}

TCP J05074264+2447555 was a microlensing event discovered toward the
Galactic anti-center \citep[$\ell = 178.7$ deg;][]{Jayasinghe17} in
data from the ASAS-SN survey \citep{Shappee14, Kochanek17}. Intensive ground-based follow-up observatories revealed a planetary signal
\citep{Nucita18}. This demonstrates the potential of an
all-sky transient survey to find microlensing events that can be
followed up with other telescopes to search for planetary
signals. Indeed, survey+follow-up strategies \citep{Gould92} have been in
place for two decades to balance the need to
monitor large numbers of stars to find microlensing events with the
need for high-cadence ($>$ few/day) observations to find planets in
those events.

\citet{Gould13LSST} showed that {\em LSST} could
conduct a microlensing survey on an unprecedented scale that covers $>
2000$ square degrees of the Galactic plane. He estimates that {\em
  LSST} could find $\sim 250$ Galactic plane microlensing events per
year, 10\% of which would be high-magnification \citep[i.e., have an
  extremely high detection efficiency for finding
  planets;][]{GriestSafizadeh98}. If real-time microlensing alerts and
targeted follow-up observations from 1--2-m class telescopes were
combined with such a survey, these high-magnification events would
give a likely return of $\sim8$--$10$ planets per year
\citep{Gould10}. In addition, follow-up observations (albeit on a much
longer timescale) could target the few events with bright, giant
sources, which are the most likely events to reveal the lower
mass-ratio planets \citep{Gould97Hollywood,
  Beaulieu06, Hwang18}. The Galactic distribution of planets as a function of $(\ell, b)$ can be
inferred by inferred from the measured planet frequencies along different
lines of sight.

The primary requirement for an {\em LSST} microlensing survey of the
Galactic plane is that instead of systematically avoiding the Plane,
it surveys the Plane at a cadence of once every 3 to 4 days, i.e., the
same cadence as for other fields. This cadence would allow sufficient
time to discover, alert, and predict high-magnification microlensing
events, so that they could be followed up with other facilities, as is
planned for other transients discovered by {\em LSST}. A higher cadence would allow earlier detection and more robust prediction of microlensing events. In addition, a
network of 1-2 m telescopes \citep[similar to Las Cumbres Observatory;
][]{Brown13} would be needed to follow up the most promising
events.

\subsection{NIR Microlensing in the Galactic Center}

The {\em UKIRT} microlensing survey\footnote{{\em UKIRT} survey light curves are publicly available through the NASA Exoplanet Archive.} has demonstrated that operating at
NIR wavelengths permits the discovery of planets close to the Galactic
center. Figure \ref{fig:galdist} shows that the first such planet
\citep[UKIRT-2017-BLG-001Lb;][]{Shvartzvald18} is deep into the
regions too extincted to be accessed by optical
surveys. \citet{Gould95Kband} shows that lenses close to the
Galactic Center are especially well characterized since their
timescales can be related directly to their distances. Furthermore,
probing lines of sight closer to the plane could reveal variations in
planet populations that parallel variations in the stellar populations.

Observing microlensing events at $b \sim 0$ requires NIR observations
because of the high extinction, preferably in $K$-band. An optimal NIR
survey to find planets would fulfill all of the usual microlensing
requirements: wide field-of-view camera, cadence of $\gtrsim 1\, {\rm hr}^{-1}$, in the Southern hemisphere, and preferably with multiple
sites. Although {\em UKIRT} does not meet all of these requirements,
it has proven to be a capable microlensing survey telescope
\citep{Shvartzvald17, Shvartzvald18}. If more time were devoted to
such a survey, it could achieve the necessary cadence to discover
Neptune mass-ratio planets ($1\, {\rm hr}^{-1}$). In addition, Prime focus Infrared Microlensing Experiment ({\em PRIME}) is a joint Japan-U.S.-South Africa NIR telescope being built in South Africa. It will have a 1.3 deg$^2$ FOV and should see
   first light in 2020. A simultaneous {\em UKIRT} survey can complement {\em PRIME} in temporal coverage,
but {\em UKIRT} is also unique in having K-band observations, which
is necessary probe all the way to the Galactic center. A survey with {\em VISTA} would also improve temporal coverage.

\section{The Planet Mass Function}

\begin{figure}
\includegraphics[height=2.5in]{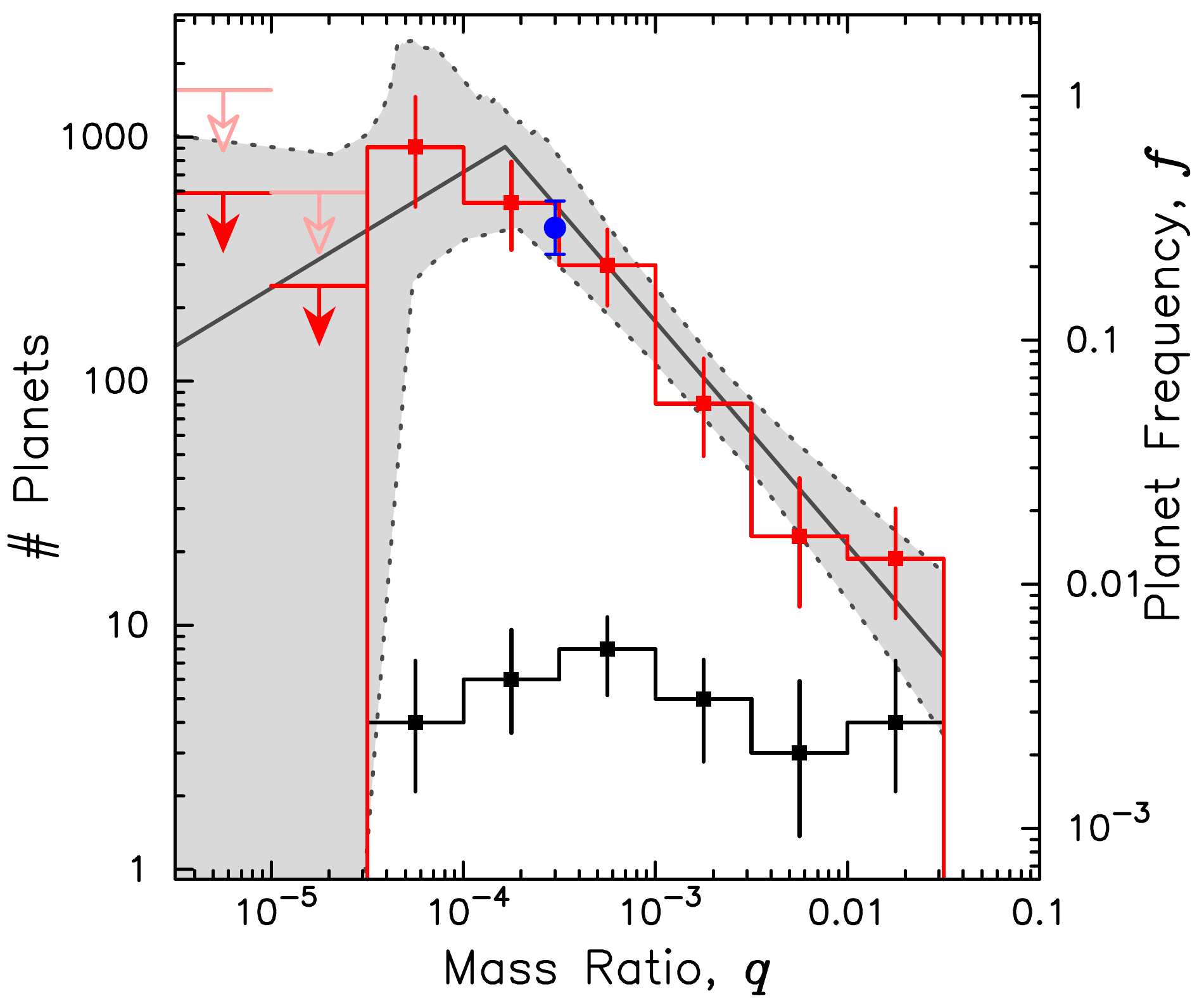}
\includegraphics[height=2.5in]{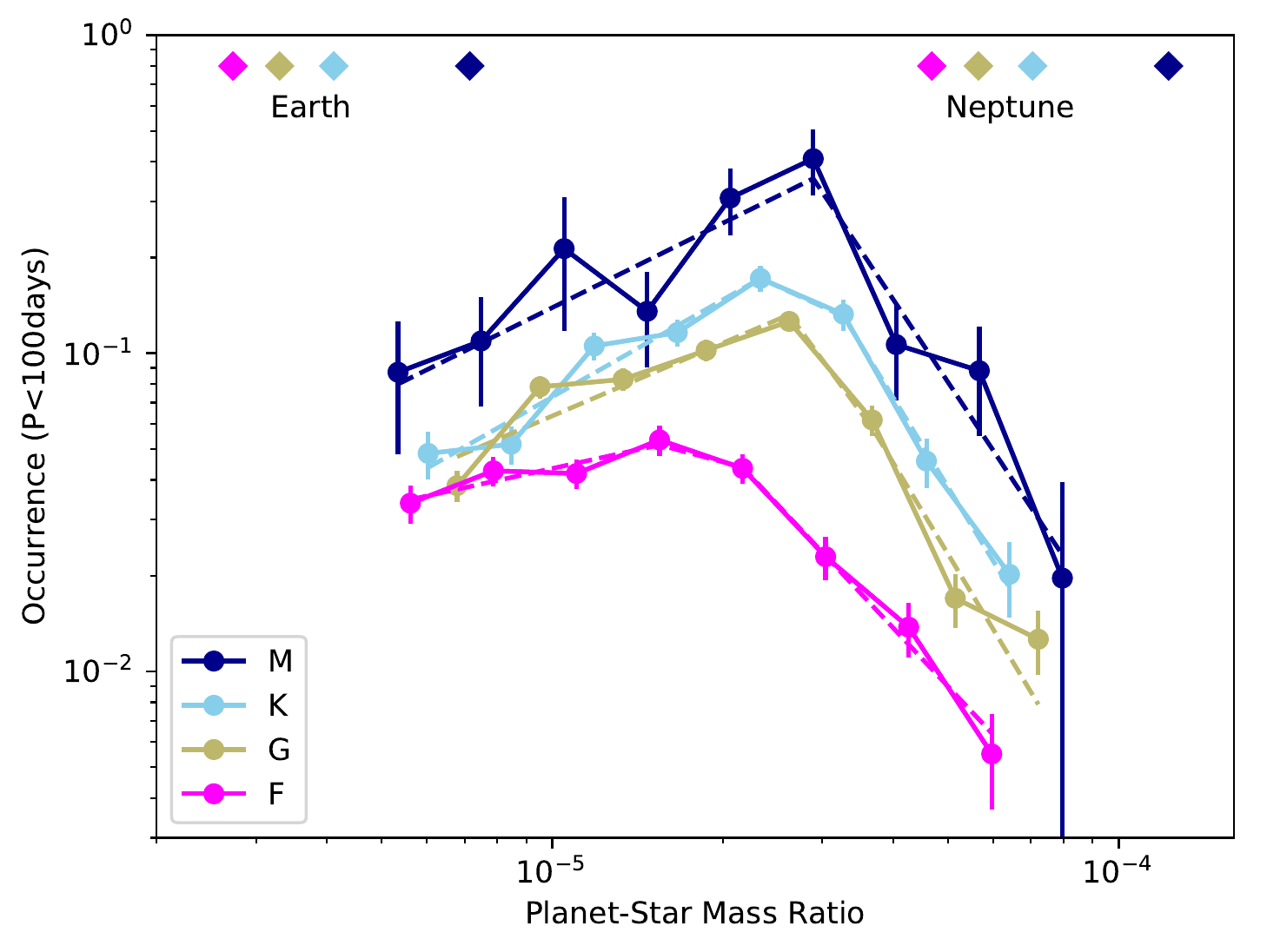}
\caption{Left: The planet mass-ratio distribution measured by
  \citet{Suzuki16} shows a break at mass-ratios $q\sim 2\times
  10^{-4}$, but the lack of planet discoveries at $q < 4\times 10^{-5}$
  leads to a large uncertainty in the mass-ratio function below the
   break. Right: \citet{Pascucci18} also see a break
  in the mass-ratio function for Kepler planets with $P<100$ days but
  at $q\sim 3\times 10^{-5}$ and find that it is independent of host
  mass. Over the next decade, ground-based microlensing surveys will
  precisely determine the mass ratio at which this break occurs for
  planets beyond the snow line. With ELTs, we can transform this into a mass distribution, allowing us to assess
  implications for planet formation theory. \label{fig:massfn}}
\end{figure}

Recent results from \citet{Suzuki16} and \citet{Udalski18} show a peak
in the mass-ratio distribution at $q = m_p/M_\star \sim 2\times
10^{-4}$ and suggest that the frequency of planets falls off
rapidly toward lower mass ratios. However, the precise mass ratio at
which this peak occurs is poorly constrained from the current
observations, and the nature of the drop-off toward lower mass ratios
is barely probed. Figure \ref{fig:massfn} shows the mass-ratio
distribution (and its large uncertainties at small mass ratios) from
\citet{Suzuki16}. If current ground-based microlensing surveys are
continued over the next decade, they will precisely measure the
location of the break in the mass-ratio distribution. Constraining the
behavior of the low mass-ratio end of the distribution requires going
to space \citep{Bennett18WP}.

The mass ratio at which this break occurs must be explained by planet formation theories, and the more precise the measurement, the stronger the constraint on the physics of planet formation. Better yet, would be to
measure the absolute masses and separations of the planets. This would
enable us to test how planet formation depends on various
factors. Some key questions that can be addressed through absolute
mass measurements are
\begin{itemize}
    \item{Is the planet mass-ratio function more fundamental than the
      planet mass function (as suggested by \citealt{Pascucci18} and
      \citealt{Udalski18})? And, the closely related question, does planet
      formation depend on stellar mass and if so, in what way?}
    \item{How does the mass-ratio function vary with separation from
      the star? Specifically, how does the $q\sim 2\times10^{-4}$ break in
      the mass-ratio function seen by microlensing at snow-line
      distances relate to the break in the mass-ratio function seen by
      Kepler at $q\sim 3\times 10^{-5}$ for $P < 100$ days
      \citep{Pascucci18}?}
    \item{How do microlensing results (which are biased toward M dwarf
      host stars) compare to results from radial velocity and transits
      (which are biased toward FGK dwarfs)? Absolute mass measurements
      would enable direct comparisons to the results from these other
      techniques by allowing the sample to be divided by host star
      mass.}
\end{itemize}
{\em WFIRST} will address these questions by measuring masses for many of
its host stars. However, a high-resolution follow-up imaging campaign from the ground would
also allow mass measurements for planets discovered by ground-based
microlensing surveys over a much larger range of environments.

Flux measurements of the lens stars (well after the microlensing
event) can be combined with measurements of the angular Einstein ring
radius (from the microlensing event) to yield masses of the host stars
(and thus planets) and distances to the systems. In addition, AO observations can be used to rule out host stars for
FFP candidates discovered by microlensing
\citep[e.g., ][]{Mroz18}. AO constraints on flux from a host can be significantly enhanced by measurement of the IR flux of the source star to a precision of $\sim 1$\%, which requires $\sim 1\,{\rm day}^{-1}$ NIR surveys simultaneous with optical surveys.

The primary challenge for making a flux measurement of the lens star
is one of resolution. By definition, the microlensing lens and source
stars are superposed to within $\sim 1$ mas at the time of the lensing
event. Furthermore, the bulge microlensing fields are extremely
crowded, so it is not uncommon to have one or more unrelated stars
blended into the PSF of the event at $\sim 1^{\prime\prime}$
seeing. However, a single, AO image taken several years
after the event can both resolve the source and lens stars as well as
unrelated, non-coincident stars. Such measurements have already been
performed in several cases \citep[e.g., ][]{Bennett10b, Batista14,
  Beaulieu18}, but, because of the resolution limit of current
facilities, these measurements are restricted to events with high proper
motions and/or those that occurred $\gtrsim 10$ years ago. In
contrast, the next generation of extremely large telescopes (e.g., TMT
and GMT), by virtue of their large diameters, could measure the fluxes
for all (or nearly all) microlensing planets within 5 years of their
discovery.

The efficiency of such a program would be vastly improved by enabling
a ``less-than-perfect fast" mode for AO observations with
30-meter class telescopes. AO systems are being
designed with the aim of achieving deep imaging with maximum contrast
and optimal Strehl ratios for the purpose of directly detecting
exoplanets. By contrast, the program proposed here is simply trying to
resolve two stars. Thus, it is not necessary to achieve either optimal
Strehl or deep imaging (a 5--10 min exposure is sufficient), which
means that the overheads in slew, settle, and acquisition time are the
limiting factors to carrying out a campaign to systematically image
all microlensing planet hosts. A little forethought to allow faster
observations (allowing for suboptimal performance) with future
AO systems would vastly improve the scientific yield of
ground-based microlensing surveys at relatively modest cost.

\bibliography{references}

\end{document}